\documentclass[twocolumn,showpacs,preprintnumbers,amsmath,amssymb,aps,floatfix,prl,superscriptaddress, nofootinbib]{revtex4-1}
\usepackage{amsfonts}
\usepackage{dcolumn}
\usepackage{bm}
\usepackage{color}
\usepackage{epsfig}
\usepackage{hyperref}

\begin{document}

\def\red#1{{\color{red} #1}}
\def\blue#1{{\color{blue} #1}}

\def\imagi{\mathrm{i}}
\def\diff{\mathrm{d}}

\def\sx{\hat{\sigma}_x}
\def\sz{\hat{\sigma}_z}
\def\q{\hat{q}}
\def\a{\hat{a}}
\def\adag{\hat{a}^{\dagger}}
\def\A{\hat{\vec{A}}}
\def\eps{\vec{\epsilon}}

\def\ket#1{\vert #1\rangle}
\def\bra#1{\langle#1\vert}
\def\braket#1#2{\langle #1 \vert #2 \rangle}

\def\br{\vec{r}}
\def\lr{\vec{\lambda}}

\def\etal{\textit{et al}}

\renewcommand\vec{\mathbf}

\title{Ground-State Quantum-Electrodynamical Density-Functional Theory}

\author{M.~Ruggenthaler}
\affiliation{Max-Planck-Institut f\"ur Strukture und Dynamik der Materie, Luruper Chaussee 149, 22761 Hamburg, Germany}
\affiliation{Institut f\"ur Theoretische Physik, Universit\"at Innsbruck, Technikerstra{\ss}e 21A, 6020 Innsbruck, Austria}

\date{\today}

\begin{abstract}
In this work we establish a density-functional reformulation of coupled matter-photon problems subject to general external electromagnetic fields and charge currents. We first show that for static minimally-coupled matter-photon systems an external electromagnetic field is equivalent to an external charge current. We employ this to show that scalar external potentials and transversal external charge currents are in a one-to-one correspondence to the expectation values of the charge density and the vector-potential of the correlated matter-photon ground state. This allows to establish a Maxwell-Kohn-Sham approach, where in conjunction with the usual single-particle Kohn-Sham equations a classical Maxwell equation has to be solved. In the magnetic mean-field limit this reduces to a current-density-functional theory that does not suffer from non-uniqueness problems and if furthermore the magnetic field is zero recovers standard density-functional theory.  
\end{abstract}
\maketitle


In the recent years tremendous experimental advances~\cite{you-2011, schwartz-2011, orgiu-2015, shalabney-2015a, shalabney-2015b, chikkaraddy-2016} have been made in the studies of complex multi-particle systems such as atoms, molecules and solid-state systems that strongly couple to the photon field~\cite{ebbesen-2016, sukharev-2017}. At the interface between quantum chemistry, solid-state physics and quantum optics interesting physical phenomena arise such as the possibility to tune chemical properties of complex systems by hybridizing them with the photon field~\cite{hutchison-2012, imorral-2012} or to generate attractive photons by coupling them strongly to matter~\cite{firstenberg-2013}. In such situations the matter and electromagnetic degrees of freedom get strongly mixed and a simple decoupling between quantized light and matter is no longer possible. Since the standard \textit{ab-initio} approaches of many-body theory~\cite{fetter-walecka,dreizler-gross,bonitz} are only concerned with the matter degrees of freedom after decoupling from the photon field, alternative \textit{ab initio} methods that treat photons and matter on equal quantized footing need to be developed in order to be applicable to such challenging physical situations. At the same time, even after decoupling, standard \textit{ab-initio} methods such as ground-state density-functional theory (DFT)~\cite{dreizler-gross, engel-dreizler} are not straightforwardly applicable to situations with strong classical electromagnetic fields. The extension of ground-state DFT to external magnetic fields, so-called current-DFT (CDFT)~\cite{vignale-1987}, is plagued by non-uniqueness problems~\cite{capelle-2002,laestadius-2014}, which hampered the development of this many-body method. With this in mind it would be highly desirable to find a density-functional formulation of the fully coupled matter-photon problem in equilibrium that leads in the magnetic mean-field limit, i.e., the photon field is treated purely classically, to a DFT for classical magnetic fields and reduces to standard DFT in the limit of no magnetic field.

In this work we provide such a density-functional formulation which we call in accordance to the non-equilibrium theory~\cite{rajagopal-1994, ruggenthaler-2011, tokatly-2013, ruggenthaler-2014, flick-2015} ground-state quantum-electrodynamical DFT (QEDFT). We do so by first showing that for the eigenstates of a general Hamiltonian in the non-relativistic limit of quantum-electrodynamics (QED) an external magnetic field is unitarily equivalent to an external transversal charge current. Therefore only an external scalar potential $v(\br)$ and an external transversal charge current $\vec{j}(\br)$ is needed to uniquely determine the Hamiltonian. Then we extend the Hohenberg-Kohn proof to show that for every such external pair $(v(\br), \vec{j}(\br))$ there is a unique ground state and that these external variables are one-to-one with the internal pair $(n(\br), \vec{A}(\br))$, where $n(\br)$ is the charge-density expectation value and $\vec{A}(\br)$ is the vector-potential expectation value of the coupled matter-photon ground-state. This makes the equilibrium situations conceptually very different from the non-equilibrium case, where the internal pair is the physical charge current and vector potential expectation values. We then introduce the corresponding Kohn-Sham scheme based on the Maxwell-Pauli Hamiltonian and discuss consequences. 

In the following we focus on the non-relativistic limit of QED~\cite{craig-1984, spohn-2004}, which should be sufficient to describe most atomic, molecular and solid-state systems and captures effects missing in pure quantum mechanics such as finite-life times of excited states, Casimir-Polder and retarded van-der-Waals forces, and polarization and local-field effects, to name but a few. The ground-state QEDFT formulation can be easily extended to semi-relativistic, i.e., including higher-order relativistic corrections, or to a fully relativistic treatment in terms of the Dirac equation for the charged particles. For the photon field and its observables we follow the standard procedure~\cite{greiner-reinhardt} and quantize in an arbitrarily large but finite box of length $L$, leading to the allowed wave vectors of the photon field $\vec{k}_{\vec{n}} = 2 \pi \vec{n}/L$. If we further make the choice of the Coulomb gauge for the photon field, i.e., the vector-potential operator should obey $\vec{\nabla} \cdot \A(\br) = 0$, then this operator reads explicitly~\cite{ruggenthaler-2014}
\begin{align*}
 \A(\br) = \sqrt{\frac{\hbar c^2}{\epsilon_0 L^3}}\sum_{\vec{n},\lambda} \frac{\vec{\epsilon}_{\vec{n},\lambda}}{\sqrt{2 \omega_{\vec{n}}}} \left(\a_{\vec{n},\lambda} e^{\imagi \vec{k}_{\vec{n}}\cdot \vec{r}} + \adag_{\vec{n},\lambda} e^{-\imagi \vec{k}_{\vec{n}}\cdot \vec{r}}\right),
\end{align*}
where $\epsilon_0$ is the vacuum permittivity, we used the usual creation $\adag_{\vec{n},\lambda}$ and annihilation $\a_{\vec{n},\lambda}$ operators for the photon frequencies $\omega_{\vec{n}} = c |\vec{k}_{\vec{n}}|$ and the polarization direction $\lambda \in \{ 1, 2 \}$, and the transversal polarization vectors obey $\vec{\epsilon}_{\vec{n},\lambda} \cdot \vec{\epsilon}_{\vec{n},\lambda'} = \delta_{\lambda,\lambda'}$ and $\vec{\epsilon}_{\vec{n},\lambda} \cdot \vec{n} =0$. 
The energy of the free photon field takes the simple form~\cite{greiner-reinhardt, ruggenthaler-2014} $\sum_{\vec{n},\lambda} \hbar \omega_n \adag_{\vec{n},\lambda} \a_{\vec{n},\lambda}$ and the energy due to coupling to a classical (external) transversal charge current
\begin{align}
\label{ExternalCurrent}
 \vec{j}(\br)\! =\! \sqrt{\frac{\epsilon_0 \hbar}{L^3}}\sum_{\vec{n},\lambda} \omega_{\vec{n}} \frac{\vec{\epsilon}_{\vec{n},\lambda}}{\sqrt{2 \omega_{\vec{n}}}} \left( j_{\vec{n},\lambda} e^{\imagi \vec{k}_{\vec{n}}\cdot \vec{r}}\! +\!  j_{\vec{n},\lambda}^{*} e^{-\imagi \vec{k}_{\vec{n}}\cdot \vec{r}}\right),
\end{align}
with the expansion coefficients $j_{\vec{n},\lambda} = j_{-\vec{n},\lambda}^{*} = (2 \omega_{\vec{n}}^3 \epsilon_0 \hbar L^3)^{-1/2} \int \diff^3 r \vec{\epsilon}_{\vec{n}, \lambda}\cdot \vec{j}(\br) \exp(- \imagi \vec{k}_{\vec{n}}\cdot \br)$, is given by the minimal coupling prescription~\cite{greiner-reinhardt, ruggenthaler-2014} as
\begin{align}
 \!\!\!-\frac{1}{c} \! \int \! \diff ^3 r  \; \vec{j}(\br) \!\cdot\! \A(\br) \! = \!\sum_{\vec{n},\lambda} \!\!-\hbar \omega_{\vec{n}}\left(   \a_{\vec{n},\lambda} j_{\vec{n},\lambda}^{*} \!+ \! j_{\vec{n},\lambda} \adag_{\vec{n},\lambda} \right).   
\end{align}
Next we couple the photon field to the charged quantum particles. For simplicity we only consider $N$ electrons, but an extension to different particles, e.g., effective quantum nuclei, is straightforward. Following the minimal coupling prescription, also allowing a classical external Coulomb-gauged vector potential
\begin{align}
 \label{ExternalPotential}
\vec{b}(\br) \!=\! \sqrt{\frac{\hbar c^2}{\epsilon_0 L^3}}\sum_{\vec{n},\lambda} \frac{\vec{\epsilon}_{\vec{n},\lambda}}{\sqrt{2 \omega_{\vec{n}}}} \left( b_{\vec{n},\lambda} e^{\imagi \vec{k}_{\vec{n}}\cdot \vec{r}} \!+\!  b_{\vec{n},\lambda}^{*} e^{-\imagi \vec{k}_{\vec{n}}\cdot \vec{r}}\right),
\end{align}
which corresponds to a static magnetic field $c \vec{B}(\br)=\vec{\nabla}\times \vec{b}(\br)$, and a scalar potential $v(\br)$, we arrive at a generalization of the so-called Pauli-Fierz Hamiltonian\footnote{We point out that in general a form-factor for the photon modes has to be employed~\cite{spohn-2004}. In our case we could choose an arbitrarily high but finite frequency as ultraviolet cutoff, and also employ the same cutoff to the external vector potentials and charge currents.}~\cite{spohn-2004, ruggenthaler-2014}
\begin{align}
\label{Hamiltonian}
 \hat{H} \!=& \!\sum_{k=1}^{N} \!\left\{ \!\frac{1}{2m}\!\left[\boldsymbol{\sigma}_{k}\!\cdot\! \left(- \imagi \hbar \vec{\nabla}_k  + \frac{|e|}{c}\left(\A(\br_k) + \vec{b}(\br_k)\right) \right) \right]^2 \!\!\! \right. \nonumber
 \\
 & \left. -|e| v(\br_i)  \right\}  
 \!+ \!\sum_{k>l}^{N}\!\!  w(|\br_k -\br_l|) \!
 \\
 &+\! \sum_{\vec{n},\lambda} \!\hbar \omega_{\vec{n}} \left\{\adag_{\vec{n},\lambda} \a_{\vec{n},\lambda} \! -  \! \a_{\vec{n},\lambda} j_{\vec{n},\lambda}^{*} \!- \! \adag_{\vec{n},\lambda} j_{\vec{n},\lambda} \right\}, \nonumber 
\end{align}
where $\boldsymbol{\sigma}$ is a vector of the usual Pauli matrices due to the spin of the electrons, $|e|$ is the elementary charge and
\begin{align*}
 w(|\br -\br'|) = \sum_{\vec{n}} \frac{e^2}{\vec{k}_{\vec{n}}^2} \frac{e^{\imagi \vec{k}_{\vec{n}}\cdot(\br-\br')}}{\epsilon_0 L^3} \overset{L\rightarrow \infty}{\longrightarrow} \frac{e^2}{4 \pi \epsilon_0 |\br-\br'|}
\end{align*}
is the interaction due to the longitudinal part of the photon field~\cite{greiner-reinhardt, ruggenthaler-2014}. The physical charge-current operator of the Pauli-Fierz Hamiltonian that obeys the continuity equation is 
\begin{align*}
 \hat{\vec{J}}(\br) = \hat{\vec{j}}_{\rm p}(\br) + \hat{\vec{j}}_{\rm m}(\br) - \frac{|e|}{m c} \hat{n}(\br) \left(\A(\br) + \vec{b}(\br) \right),
\end{align*}
where the first term is the paramagnetic current
\begin{align*}
 \hat{\vec{j}}_{\rm p}(\br) = -\frac{|e| \hbar}{2 m \imagi}\sum_{k=1}^{N} \left(\delta^3(\br -\br_k)\overrightarrow{\nabla}_k - \overleftarrow{\nabla}_{k} \delta^3(\br-\br_k)  \right),
\end{align*}
the second term is the magnetization current
\begin{align*}
 \hat{\vec{j}}_{\rm m}(\br) = \frac{|e|\hbar}{2 m} \sum_{k=1}^{N} \vec{\nabla}_k \times \left( \boldsymbol{\sigma}_k \delta^{3}(\br-\br_k) \right),
\end{align*}
and the third term is the diamagnetic current that also contains the photon and the external field with the charge density operator $\hat{n}(\br) = -|e| \sum_{k} \delta^{3}(\br - \br_k)$. Further, using the Heisenberg picture of quantum mechanics~\cite{greiner-reinhardt} (indicated by subindex $\rm H$) one can show that the vector-potential operator obeys the inhomogeneous Maxwell equation
\begin{align}
\label{InhomogeneousMaxwell}
 \left(\frac{1}{c^2}\frac{\diff^2}{\diff t^2} - \vec{\nabla}^2 \right) \A_{\rm H}(\br,t) = \mu_{0}c \left(\hat{\vec{J}}_{{\rm H}, \perp}(\br,t) + \vec{j}(\br)  \right),
\end{align}
where $\mu_0 = 1/(\epsilon_0 c^2)$ is the vacuum permeability and $\hat{\vec{J}}_{\perp}(\br)$ is the transversal part of the charge-current operator~\cite{ruggenthaler-2014}.

If considering the general Hamiltonian of Eq.~\ref{Hamiltonian} together with the mode-resolved expressions of the external fields in Eqs.~\ref{ExternalCurrent} and~\ref{ExternalPotential}, by applying the unitary displacement operator
\begin{align*}
 \hat{D}[b] = \exp\left(\sum_{\vec{n}, \lambda}( b_{\vec{n},\lambda} \a_{\vec{n},\lambda} - b_{\vec{n},\lambda}^{*} \adag_{\vec{n},\lambda} )\right),
\end{align*}
which just shifts $\a_{\vec{n},\lambda} \rightarrow \a_{\vec{n},\lambda} - b_{\vec{n},\lambda}
$, $\adag_{\vec{n},\lambda} \rightarrow \adag_{\vec{n},\lambda} - b_{\vec{n},\lambda}^{*}$
and thus $\A(\br) \rightarrow \A(\br) - \vec{b}(\br)$, the above Hamiltonian takes on an unitarily equivalent form
\begin{align}
\label{TransformedHamiltonian}
 \hat{H}^{\rm D} &= \hat{D}[b] \hat{H} \hat{D}^{\dagger}[b] 
 \\
 = &\sum_{k=1}^{N} \left\{ \frac{1}{2m}\left[\boldsymbol{\sigma}_{k}\cdot \left(- \imagi \hbar \vec{\nabla}_k  + \frac{|e|}{c}\A(\br_k) \right) \right]^2 -  |e| v(\br_i)  \right\}  \nonumber
 \\
 &+ \sum_{k>l}^{N}  w(|\br_k -\br_l|) + \sum_{\vec{n},\lambda} \hbar \omega_{\vec{n}} \adag_{\vec{n},\lambda} \a_{\vec{n},\lambda} \nonumber
 \\
 & - \sum_{\vec{n},\lambda} \hbar \omega_{\vec{n}} \left\{\a_{\vec{n},\lambda} \left(j_{\vec{n},\lambda}^{*} + b_{\vec{n},\lambda}^{*} \right) \!+ \! \adag_{\vec{n},\lambda} \left(j_{\vec{n},\lambda} +  b_{\vec{n},\lambda}\right) \right\}  \nonumber
 \\
 &+ \sum_{\vec{n},\lambda} \hbar \omega_{\vec{n}} \left\{ |b_{\vec{n},\lambda}|^2 + b_{\vec{n},\lambda}j_{\vec{n},\lambda}^{*} + b_{\vec{n},\lambda}^{*} j_{\vec{n},\lambda} \right\}, \nonumber
\end{align}
where the last term is just a constant shift that can be disregarded. We have thus removed the external vector potential by recasting it in terms of an external charge current. By construction, the eigenstates of the original Hamiltonian $\Psi_i$ are connected to the eigenstates of the new Hamiltonian by
\begin{align*}
\Psi_i^{\rm D} = \hat{D}[b] \Psi_i.
\end{align*}
Consequently, we can always solve instead of a problem with an external vector potential $\vec{b}(\br)$ an equivalent problem with an additional external current and determine all physical observables from $\Psi_i^{\rm D}$.  Specifically, all observables of the matter subsystem and even the physical charge current that depends on the photon field stay invariant. The latter is because the diamagnetic term for the physical current of $\hat{H}^{\rm D}$ does no longer contain an external potential term, but upon evaluation becomes
\begin{align*}
 \braket{\Psi_i^{\rm D}}{\hat{n}(\br)\A(\br) \Psi_i^{\rm D}} = \braket{\Psi_i}{\hat{n}(\br)\left(\A(\br) + \vec{b}(\br)\right) \Psi_i},
\end{align*}
and thus $\vec{J}(\br) = \vec{J}^{\rm D}(\br)$. Not surprisingly, having an external classical current or an external classical potential does not make a physical difference, at least for static problems, since we can consider $\vec{b}(\br)$ physically equivalent to $\vec{j}(\br)$ by the Maxwell relations
\begin{align*}
 - \vec{\nabla}^2 \vec{b}(\br) = \mu_0 c \vec{j}(\br).
\end{align*}
We now use this equivalence relation to establish a density-functional reformulation of time-independent non-relativistic QED. Although we could also recast the external scalar potential $v(\br)$ in terms of an external charge density, it is beneficial for the following considerations to keep the scalar potential in analogy to standard electronic ground-state DFT. 

All possible physically inequivalent Pauli-Fierz Hamiltonians can be labeled by their respective external pair $(v(\br), \vec{j}(\br))$, i.e., we write $\hat{H}[v,\vec{j}]$ for the Hamiltonian of Eq.~\ref{Hamiltonian} with $\vec{b}(\br) = 0$. We note that we have fixed the gauge freedom in the charge current and thus in the external vector potential by assuming them transversal, i.e., in Coulomb gauge, and we fix the arbitrary gauge constant in the external potential to zero. To keep the derivations simple, we follow the original formulation of Hohenberg and Kohn and restrict to non-degenerate ground states. The extension to degenerate ones is straightforward and follows the standard text-books derivations~\cite{dreizler-gross, engel-dreizler}. The first step is to show that there is one and only one ground state for a given external pair, i.e., $(v(\br), \vec{j}(\br)) \overset{1:1}{\leftrightarrow} \Psi_0$. We do so by \textit{reductio ad absurdum} and show that the opposite assumption, i.e., that $\Psi_0 = \Psi_0'$ for different Hamiltonians $\hat{H}[v,\vec{j}]$ and $\hat{H}[v',\vec{j}']$, leads to a contradiction. We first note that due to the time-independent inhomogeneous Maxwell equation
\begin{align*}
 - \vec{\nabla}^2 \vec{A}(\br) = \mu_0 c\left(\vec{J}_{\perp}(\br) + \vec{j}(\br) \right)
\end{align*}
that relates the expectation values of $\A(\br)$ and $\hat{\vec{J}}(\br)$ we necessarily have $\vec{j}(\br) = \vec{j}'(\br)$. By then subtracting $\hat{H}[v',\vec{j}] \Psi_0 = E_0' \Psi_0$ from $\hat{H}[v,\vec{j}] \Psi_0 = E_0 \Psi_0$ we find
\begin{align*}
 \sum_{k=1}^{N} - |e| \left(v(\br_k) - v'(\br_k)  \right) \Psi_0 = \left(E_0 - E_0' \right)\Psi_0,
\end{align*}
which leads to the condition that $v(\br) - v'(\br) = c$, which is a contradiction to our original assumption.

Next we show that also $\Psi_0 \overset{1:1}{\leftrightarrow} (n(\br), \vec{A}(\br))$. Again we proceed by assuming the opposite, i.e., while $\Psi_0 \neq \Psi_0'$ they have the same internal pair $\braket{\Psi_0}{\hat{n}(\br)\Psi_0} = \braket{\Psi_0'}{\hat{n}(\br)\Psi_0'}$ and $\braket{\Psi_0}{\A(\br)\Psi_0}= \braket{\Psi_0'}{\A(\br)\Psi_0'}$. With the definition of
\begin{align*}
 \hat{H}_0 &= \sum_{k=1}^{N} \left\{ \frac{1}{2m}\left[\boldsymbol{\sigma}_{k}\cdot \left(- \imagi \hbar \vec{\nabla}_k  + \frac{|e|}{c} \A(\br_k)\right) \right]^2 \right\}  
  \\
  &+ \sum_{k>l}^{N}  w(|\br_k -\br_l|) + \sum_{\vec{n},\lambda} \hbar \omega_{\vec{n}} \adag_{\vec{n},\lambda} \a_{\vec{n},\lambda}, 
\end{align*}
and the notation $\braket{\Psi_0}{\a_{\vec{n}, \lambda} \Psi_0}=a_{\vec{n}, \lambda}$ and $\tilde{a}_{\vec{n}, \lambda} = (a_{\vec{n}, \lambda} + a_{-\vec{n}, \lambda}^{*})$ we find
\begin{align*}
 E_0 = \braket{\Psi_0}{\hat{H}_0 \Psi_0} + \int \diff^3 r \; n(\br) v(\br) - \sum_{\vec{n}, \lambda} \hbar \omega_{\vec{n}} \tilde{a}_{\vec{n}, \lambda} j_{\vec{n},\lambda}^{*}
\end{align*}
and accordingly for the primed system. With the help of the previous result, i.e., that $\Psi_0 \neq \Psi_0'$ implies $(v(\br),\vec{j}(\br)) \neq (v'(\br), \vec{j}'(\br))$, this allows us find the following inequality for $E_0$ in terms of $E_0'$
\begin{align*}
 E_0 &< \braket{\Psi_0'}{\hat{H}[v',\vec{j}'] \Psi_0'} + \int \diff^3 r \; n(\br) \left(v(\br) -v'(\br) \right) \nonumber
 \\
 &- \sum_{\vec{n}, \lambda} \hbar \omega_{\vec{n}} \tilde{a}_{\vec{n}, \lambda} \left(j_{\vec{n},\lambda}^{*} - j_{\vec{n},\lambda}^{'*} \right), 
\end{align*}
and an according one for $E_0'$ in terms of $E_0$. Adding both inequalities leads to $E_0' + E_0 > E_0' + E_0$ which is clearly a contradiction.

With this we have established a mapping $(v(\br), \vec{j}(\br))  \overset{1:1}{\leftrightarrow} (n(\br), \vec{A}(\br))$ and all ground-state wave functions can be uniquely labeled by their respective internal pair $\Psi_0[n,\vec{A}]$. Thus, by defining a generalized universal functional\footnote{For simplicity, we only use the Hohenberg-Kohn-form of the universal functional. The Levy-Lieb constrained-search and the Lieb form of the universal functional are defined in a straightforward manner~\cite{engel-dreizler}.}
\begin{align*}
 F[n,\vec{A}] = \braket{\Psi[n,\vec{A}]}{\hat{H}_0 \Psi_0[n, \vec{A}]}
\end{align*}
we can find the ground-state of a general non-relativistic QED problem by
\begin{align*}
E[v, \vec{j}] \!\! = \!\!\inf_{(n,\vec{A})} \left\{\! F[n,\vec{A}] \!+\!\! \int \!\diff^3 r n(\br) v(\br) \!-\! \frac{1}{c}\! \int\! \diff^3 r \vec{A}(\br)\! \cdot\! \vec{j}(\br)\!\right\}.
\end{align*}
The minimum can equivalently be found by functional variation
\begin{align*}
 \frac{\delta F[n, \vec{A}]}{\delta n(\br)} = - v(\br), \qquad \frac{\delta F[n, \vec{A}]}{\delta \vec{A}(\br)} = \frac{1}{c}\vec{j}(\br).
\end{align*}
To make QEDFT practical we follow the usual way of defining a numerically simpler auxiliary system. In our case we use the Maxwell-Pauli Hamiltonian 
\begin{align*}
 \hat{H}_{\rm s} = \hat{H}_{0}^{\rm s} + \sum_{k=1}^{n}-|e|v_{s}(\br) - \frac{1}{c} \int \diff^3 r \vec{A}(\br)\cdot \vec{j}_s(\br)
\end{align*}
with the universal part
\begin{align}
\label{UniversalMaxwellPauli}
 \hat{H}_0^{\rm s} &= \sum_{k=1}^{N} \left\{ \frac{1}{2m}\left[\boldsymbol{\sigma}_{k}\cdot \left(- \imagi \hbar \vec{\nabla}_k  + \frac{|e|}{c} \vec{A}(\br_k)\right) \right]^2 \right\}  
 \\
 &  + \frac{1}{c}\int \diff^3 r \vec{J}_{\rm s}(\br) \cdot \vec{A}(\br) \nonumber
 \\
 & + \sum_{\vec{n},\lambda} \hbar \omega_{\vec{n}} \left\{\adag_{\vec{n},\lambda} \a_{\vec{n},\lambda}  -   \a_{\vec{n},\lambda} J_{\vec{n},\lambda}^{{\rm s},*} \!- \! \adag_{\vec{n},\lambda} J_{\vec{n},\lambda}^{\rm s} \right\}. \nonumber
\end{align}
Here $\vec{A}(\br)$  and $J_{\vec{n}, \lambda}^{\rm s}$ are just the expectation values of the corresponding (mode-resolved) operators. We thus have a mean-field coupling between the photons and the non-interacting electrons, and the second term in Eq.~\ref{UniversalMaxwellPauli} takes care to not double count the mean-field interaction energy. We can then perform exactly the same steps to show that $(v_s(\br), \vec{j}_s(\br))  \overset{1:1}{\leftrightarrow} (n(\br), \vec{A}(\br))$. The only subtlety is found in the second part of the generalized Hohenberg-Kohn theorem, where we cannot assume that even though two different wave functions have the same internal pair they also share the same physical charge current $\vec{J}_{\rm s}(\br)$ that goes into the third term of Eq.~\ref{UniversalMaxwellPauli}. But since by construction $\frac{1}{c} \int \diff^3 r \vec{A}(\br) \cdot \vec{J}_{\rm s}(\br) - \sum_{\vec{n}, \lambda} \hbar \omega_{\vec{n}} \tilde{a}_{\vec{n}, \lambda} J_{\vec{n}, \lambda}^{\rm s} = 0$ we find the same contradiction as before. This then allows us to perform a composition of maps $(v(\br), \vec{j}(\br))  \overset{1:1}{\leftrightarrow} (n(\br), \vec{A}(\br)) \overset{1:1}{\leftrightarrow} (v_s(\br), \vec{j}_s(\br)) $ and define as is usually done in time-dependent DFT~\cite{ruggenthaler-2015} the Hartree-exchange-correlation (Hxc) potential and exchange-correlation (xc) current directly by
\begin{align*}
 v_{\rm Hxc}([n, \vec{A}], \br) &=  v_s([n, \vec{A}], \br) - v([n,\vec{A}], \br),
 \\
 \vec{j}_{\rm xc}([n, \vec{A}], \br) &=  \vec{j}_s([n, \vec{A}], \br) - \vec{j}([n,\vec{A}], \br).
\end{align*}
We note that due to the fact that we made the mean-field coupling explicit in the Maxwell-Pauli equation we only have an xc contribution in the external current, but the Hxc potential still contains the usual Hartree term. Equivalently, with the auxiliary universal functional
\begin{align*}
 T[n,\vec{A}] = \braket{\Phi[n,\vec{A}]}{\hat{H}_{0}^{\rm s} \Phi[n, \vec{A}]},
\end{align*}
where $\Phi$ is a Slater determinant of orbitals $\phi_{k}(\br, \tau)$ and $\tau$ the spin coordinate, and the Hxc energy
\begin{align*}
 E_{\rm Hxc}[n, \vec{A}] = F[n,\vec{A}] - T[n, \vec{A}],  
\end{align*}
we can derive the Hxc potential and the xc current as functional derivatives of $E_{\rm Hxc}[n, \vec{A}]$. This allows us to determine the ground state properties by solving coupled non-linear single-particle Kohn-Sham equations of the form
\begin{align}
 E_k \phi_k(\br, \tau) &= \left\{\frac{1}{2m}\left[\boldsymbol{\sigma}\cdot\left(- \imagi \hbar \vec{\nabla} + \frac{|e|}{c} \vec{A}(\br)  \right)  \right]^2 + v(\br)  \right. \nonumber
 \\
 &\left. \qquad + v_{\rm Hxc} ([n, \vec{A}], \br) \right\} \phi_{k}(\br, \tau)
 \\
 E_{\vec{n}, \lambda} \varphi_{\vec{n}, \lambda}(m) &= \left\{\hbar \omega_{\vec{n}} \adag_{\vec{n},\lambda}\a_{\vec{n},\lambda} - \a_{\vec{n},\lambda}\left(j_{\vec{n},\lambda} + J^{s}_{\vec{n},\lambda} + j^{\rm xc}_{\vec{n},\lambda}\right)^{*} \right. \nonumber
 \\
 & \left. \qquad - \adag_{\vec{n},\lambda}\left(j_{\vec{n},\lambda} + J^{s}_{\vec{n},\lambda} + j^{\rm xc}_{\vec{n},\lambda}\right)  \right\} \varphi_{\vec{n}, \lambda}(m),
\end{align}
where $m$ is the mode occupation, $J_{\rm s}(\br)$ is the physical current of the Maxwell-Pauli Hamiltonian, $n(\br) = \sum_{k, \tau} - |e| |\phi_k(\br, \tau)|^2$ and $\vec{A}(\br)=\sum_{\vec{n}, \lambda}\braket{\varphi_{\vec{n}, \lambda}}{\A(\br)\varphi_{\vec{n}, \lambda}}$. Since the equations for the photon field are just a sum of shifted harmonic oscillators this can be equivalently written as a classical Maxwell equation
\begin{align}
 - \vec{\nabla}^2 \vec{A}(\br) = \mu_0 c\left(\vec{j}(\br) + \vec{J}^{\rm s}_{\perp}(\br) + \vec{j}_{\rm xc}([n,\vec{A}], \br)  \right). 
\end{align}
Thus instead of solving for the infeasible non-relativistic QED ground state one can instead solve $N$ non-linear single-particle equations coupled to a classical Maxwell equation. This is the main result of this letter.

Now, let's consider the consequences of the different results. First, there is formally no difference between an external vector potential and an external charge current in the equilibrium case. In the time-dependent case, however, external charge currents and vector potentials are formally different~\cite{ruggenthaler-2011, tokatly-2013, ruggenthaler-2014}. The physical reason is that in the time-independent case the employed unitary transformation is merely changing the vacuum of the bare photon field and with this the arbitrary reference to which we gauge our electromagnetic quantities. In the time-dependent case we have fixed by our initial state the reference and a change of the vacuum corresponds to a real physical change. This makes ground-state and time-dependent QEDFT conceptually different. Next, if we make the magnetic mean-field approximation already in the interacting Hamiltonian~\ref{Hamiltonian}, i.e., assuming a Maxwell-Pauli Hamiltonian~\ref{UniversalMaxwellPauli} with Coulomb interaction, then the resulting maps will change and we find the semi-classical limit of QEDFT. If we then compare to standard CDFT~\cite{vignale-1987} that treats matter coupled to classical magnetic fields (but the matter does not act back), we realize that it is the change of perspective to go from $(v(\br), \vec{b}(\br))$ to $(v(\br), \vec{j}(\br))$ that avoids the shortcomings of CDFT~\cite{laestadius-2014}. In this theory different external fields can have the same ground state~\cite{capelle-2002} which makes CDFT much more challenging then DFT. For instance, a Hamiltonian with a uniform magnetic field and one where we simultaneously change the scalar potential and the magnetic field can share the same electronic ground state, even though their external fields are physically different. However, this change immediately affects the physical current of the system and with this $\vec{A}(\br)$. In this way the electromagnetic degrees of freedom become a measure for the difference in external potentials. The change of perspective also changes the effective fields that are used. Instead of an xc vector potential that is employed in standard CDFT we have an xc current $\vec{j}_{\rm xc}(\br)$ that provides the missing transversal correlation effects. In the case that the coupling to the transversal photon degrees is disregarded completely while keeping the longitudinal photon degrees that lead to the Coulomb interaction, we recover standard electronic ground-state DFT. This makes QEDFT a natural generalization of DFT. Especially, since given approximations to the xc functionals, this generalization is particularly simple to implement in existing DFT codes. The vector potential is just $A(\vec{r}) = \int \diff^3 r' w(|\br-\br'|) \left(\vec{j}(\br') + \vec{J}^{\rm s}(\br') + \vec{j}_{\rm xc}([n,\vec{A}], \br') \right)/c e^2$. This then allows to calculate ground-state properties of systems coupled to external magnetic fields $\vec{b}(\br)$ and even pure quantum-field effects such as the Lamb shift or Casimir-Polder and retarded van-der-Waals forces by merely coupling non-linearly to a classical Maxwell field. The only thing missing to make QEDFT practical are the approximations to the xc functionals. For this we can rely on strategies developed for electronic DFT. Besides possibilities like a re-evaluation of the local-density approximation but now coupled to the photon field, especially the optimized-effective potential approach~\cite{kuemmel-2003,kuemmel-2008}, where energy expressions in terms of the auxiliary wave-functions are used, is straightforwardly applicable. Indeed, this has already been done successfully~\cite{pellegrini-2015, flick-2016} and first calculations for real molecules strongly-coupled to photons in an optical cavity~\cite{flick-2017} highlight the potential of QEDFT. Since these first calculations are performed in the dipole approximation, let us finally comment on this special but very important limit of matter-photon coupling. In this case we only keep a few of the photon modes which we denote by $\alpha$, and assume that the wavelengths of these modes are much larger than the 
extend of the matter subsystem. Then we approximate the mode functions by $\exp(\pm \imagi \vec{k}_{\alpha}\cdot \vec{r}) \approx 1$, which allows us to simplify the original Hamiltonian~\ref{Hamiltonian} considerably by going into the length-gauge~\cite{tokatly-2013, ruggenthaler-2014, flick-2015, flick-2016, dimitrov-2017}. Since then we only have external scalar potentials we do not need to perform a shift of the vacuum and can use the above generalized Hohenberg-Kohn approach directly to establish a QEDFT formulation for the internal pair $(n(\br),\{ q_{\alpha}\})$, where  $q_{\alpha}$ corresponds to the expectation value of the displacement coordinate of the photon mode $\alpha$~\cite{tokatly-2013, ruggenthaler-2014}.

To conclude, we have established ground-state QEDFT and the corresponding Maxwell-Kohn-Sham construction. This puts \textit{ab initio} calculations for coupled matter-photon systems subject to arbitrary external electromagnetic fields and currents on firm theoretical grounds. Further developments within QEDFT include matter-photon functionals and numerical implementations beyond the dipole approximation, which will allow us to investigate local-field, polarization and retardation effects in real molecules and solid-state systems.


{\em Acknowledgement.}
I want to thank M.\ Penz, A.\ Laestadius, J.\ Flick, H.\ Appel and F.\ Eich for valuable discussions and want to acknowledge the referees that helped to improve the work considerably. Further, I acknowledge financial support by the FWF (Austrian Science Fund) through project P25739-N27.
  

\begin{thebibliography}{99}






\bibitem{schwartz-2011} T.\ Schwartz, J.A.\ Hutchison, C.\ Genet, and T.W.\ Ebbesen, \href{http://journals.aps.org/prl/abstract/10.1103/PhysRevLett.106.196405}{Phys. Rev. Lett. {\bf 106}, 196405 (2011).}

\bibitem{you-2011} J.Q.\ You, and F.\ Nori, \href{http://www.nature.com/nature/journal/v474/n7353/full/nature10122.html}{Nature {\bf 474}, 589 (2011).}

\bibitem{orgiu-2015} E.\ Orgiu \etal, \href{http://www.nature.com/nmat/journal/v14/n11/full/nmat4392.html}{Nature Materials {\bf 14}, 1123 (2015).}

\bibitem{shalabney-2015a} A.\ Shalabney \textit{et.al.}, \href{http://onlinelibrary.wiley.com/doi/10.1002/anie.201502979/full}{Angew. Chem. Int. Ed. \textbf{54}, 7971 (2015).}

\bibitem{shalabney-2015b} A.\ Shalabney \textit{et.al.}, \href{https://www.nature.com/articles/ncomms6981}{Nature Comm. \textbf{6}, 5981 (2015).}

\bibitem{chikkaraddy-2016} R.\ Chikkaraddy \textit{et.al.}, \href{https://www.nature.com/nature/journal/v535/n7610/full/nature17974.html}{Nature \textbf{535}, 127 (2016).}

\bibitem{ebbesen-2016} T.\ Ebbesen, \href{http://pubs.acs.org/doi/abs/10.1021/acs.accounts.6b00295}{Acc. Chem. Res. \textbf{49}, 2403 (2016).}

\bibitem{sukharev-2017} S.\ Sukharev, and A.\ Nitzan, \href{https://arxiv.org/abs/1704.05605}{preprint arXiv:1704.05605 (2017).}


\bibitem{hutchison-2012} J.A.\ Hutchison, T.\ Schwartz, C.\ Genet, E.\ Devaux, and T.W.\ Ebbesen, \href{http://dx.doi.org/10.1002/anie.201107033}{Angew. Chem. Int. Ed. {\bf 51}, 1592 (2012).}

\bibitem{imorral-2012} A.F.\ i Morral, and F.\ Stellacci, \href{http://www.nature.com/nmat/journal/v11/n4/full/nmat3284.html}{Nature Materials {\bf 11}, 272 (2012).}

\bibitem{firstenberg-2013} O.\ Firstenberg \textit{et.al.}, \href{https://www.nature.com/nature/journal/v502/n7469/abs/nature12512.html}{Nature \textbf{502}, 71 (2013).}

\bibitem{fetter-walecka} A.L.\ Fetter and J.D.\ Walecka, \textit{Quantum Theory of Many-Particle Systems} (Dover Publications, 2003).

\bibitem{bonitz} M.\ Bonitz, \textit{Quantum Kinetic Theory} (Teubner-Verlag, 1998).

\bibitem{dreizler-gross} R.M.\ Dreizler and E.K.U.\ Gross, \textit{Density Functional Theory - An Approach to the Quantum Many-Body Problem} (Springer-Verlag, 1990).

\bibitem{engel-dreizler} E.\ Engel and R.M.\ Dreizler, \textit{Density Functional Theory - An Advanced Course} (Springer-Verlag, 2011).


\bibitem{vignale-1987} G.\ Vignale, and M.\ Rasolt. \href{https://journals.aps.org/prl/abstract/10.1103/PhysRevLett.59.2360}{Phys. Rev. Lett. \textbf{59}, 2360 (1987).}

\bibitem{capelle-2002} K.\ Capelle, and G. Vignale, \href{https://journals.aps.org/prb/abstract/10.1103/PhysRevB.65.113106}{Phys. Rev. B. \textbf{65}, 113106 (2002).}

\bibitem{laestadius-2014} A.\ Laestadius, and M.\ Benedicks, \href{http://onlinelibrary.wiley.com/doi/10.1002/qua.24668/full}{Int. J. Quant. Chem. \textbf{114}, 782 (2014).}


\bibitem{rajagopal-1994} A.K. Rajagopal, \href{http://journals.aps.org/pra/abstract/10.1103/PhysRevA.50.3759}{Phys. Rev. A {\bf 50}, 3759 (1994).}

\bibitem{ruggenthaler-2011} M.\ Ruggenthaler, F.\ Mackenroth, and D.\ Bauer, \href{http://journals.aps.org/pra/abstract/10.1103/PhysRevA.84.042107}{Phys. Rev. A {\bf 84}, 042107 (2011).}

\bibitem{tokatly-2013} I.V.\ Tokatly, \href{http://journals.aps.org/prl/abstract/10.1103/PhysRevLett.110.233001}{Phys. Rev. Lett. {\bf 110}, 233001 (2013).}

\bibitem{ruggenthaler-2014} M.\ Ruggenthaler, J.\ Flick, C.\ Pellegrini, H.\ Appel, I.V.\ Tokatly, and A.\ Rubio, \href{http://journals.aps.org/pra/abstract/10.1103/PhysRevA.90.012508}{Phys. Rev. A {\bf 90}, 012508 (2014).}

\bibitem{flick-2015} J.\ Flick, \etal, \href{http://www.pnas.org/content/112/50/15285.short}{PNAS {\bf 112}, 15285 (2015).}


\bibitem{craig-1984} \href{http://store.doverpublications.com/0486402142.html}{\textit{Molecular Quantum Electrodynamics} (Academic Press, 1984).}

\bibitem{spohn-2004} H.\ Spohn, \href{http://www.cambridge.org/catalogue/catalogue.asp?isbn=9780521836975}{\textit{Dynamics of Charged Particles and Their Radiation Field} (Cambridge University Press, 2004).}


\bibitem{greiner-reinhardt} W.\ Greiner,and J.\ Reinhardt, \textit{Field quantization} (Springer, 1996).

\bibitem{ruggenthaler-2015} M.\ Ruggenthaler, M.\ Penz, and R. van Leeuwen, \href{http://iopscience.iop.org/article/10.1088/0953-8984/27/20/203202/meta}{J. Phys.: Condens. Matter \textbf{27}, 203202 (2015).}

\bibitem{kuemmel-2003} S.\ Kuemmel, and J.P.\ Perdew, \href{https://journals.aps.org/prl/abstract/10.1103/PhysRevLett.90.043004}{Phys. Rev. Lett. \textbf{90}, 043004 (2003).}

\bibitem{kuemmel-2008} S.\ Kuemmel, and L.\ Kronik, \href{https://journals.aps.org/rmp/abstract/10.1103/RevModPhys.80.3}{Rev. Mod. Phys. \textbf{80}, 3 (2008).}

\bibitem{pellegrini-2015} C.\ Pellegrini, J.\ Flick, I.V.\ Tokatly, H.\ Appel, and A.\ Rubio,
Phys. Rev. Lett. {\bf 115}, 093001 (2015).


\bibitem{flick-2016} J.\ Flick \textit{et.al.} \href{http://www.pnas.org/content/114/12/3026.short}{PNAS \textbf{114}, 3026 (2016).}


\bibitem{flick-2017} J.\ Flick \textit{et.al.} \href{}{in preparation (2017).}






\bibitem{dimitrov-2017} T.\ Dimitrov, \textit{et.al.}, \href{https://arxiv.org/abs/1706.08852}{submitted to New J. Phys. arXiv:1706.08852 (2017).}



%
%
%
%
%
%
%
%
%
%
%
%
%











\end{thebibliography}

\end{document}